%% file: main.tex
\documentclass[10pt, conference, letterpaper]{IEEEtran}

\usepackage{amsmath,amsfonts}

\usepackage{amssymb}
\usepackage{algorithmic}
\usepackage{caption}
\usepackage{subfigure}
\usepackage{graphicx}
\usepackage{textcomp}
\usepackage{xcolor}
\usepackage{url}
\usepackage{booktabs}

\usepackage{authblk}

\def\BibTeX{{\rm B\kern-.05em{\sc i\kern-.025em b}\kern-.08em
    T\kern-.1667em\lower.7ex\hbox{E}\kern-.125emX}}

\IEEEoverridecommandlockouts\IEEEpubid{\makebox[\columnwidth]{979-8-3503-9973-8/23/\$31.00 $\copyright$2023 IEEE\hfill}\hspace{\columnsep}\makebox[\columnwidth]{ }}  

\begin{document}

\title{SDT: A Low-cost and Topology-reconfigurable Testbed for Network Research
\thanks{* Corresponding author: Yang Xu.}
\thanks{This paper will be published in IEEE CLUSTER 2023. Preview version only.}
}

\input{Contents/Authors}

\maketitle

\input{Contents/Abstract}


\input{body}

\section*{Acknowledgements}
This work is sponsored by the Key-Area Research and Development Program of Guangdong Province (2021B0101400001), National Natural Science Foundation of China (62150610497, 62172108, 62002066), Natural Science Foundation of Shanghai (23ZR1404900), the Major Key Project of PCL, and Open Research Projects of Zhejiang Lab (2022QA0AB07). We also sincerely appreciate the anonymous reviewers for their valuable and constructive feedback.

\bibliographystyle{Touko-Format-unsrt}
\bibliography{main}

\end{document}

%% file: Contents/Authors.tex
\author[$\dagger$]{Zixuan Chen}
\author[$\dagger$]{Zhigao Zhao}
\author[$\dagger$]{Zijian Li}
\author[$\dagger$]{Jiang Shao}
\author[$\dagger \ddagger$]{Sen Liu}
\author[$\dagger \diamondsuit \ast$]{Yang Xu}
\affil[ ]{\{zxchen20, zgzhao20, lizj21, jshao20, senliu, xuy\}\ @fudan.edu.cn}
\affil[$\dagger$]{School of Computer Science, Fudan University, Shanghai, China}
\affil[$\ddagger$]{Institute of Fintech, Fudan University, Shanghai, China}
\affil[$\diamondsuit$]{Peng Cheng Laboratory, Shenzhen, China}


%% file: Contents/Abstract.tex
\begin{abstract}

Network experiments are essential to network-related scientific research (e.g., congestion control, QoS, network topology design, and traffic engineering). However, (re)configuring various topologies on a real testbed is expensive, time-consuming, and error-prone. In this paper, we propose \emph{Software Defined Topology Testbed (SDT)}, a method for constructing a user-defined network topology using a few commodity switches. SDT is low-cost, deployment-friendly, and reconfigurable, which can run multiple sets of experiments under different topologies by simply using different topology configuration files at the controller we designed. We implement a prototype of SDT and conduct numerous experiments. Evaluations show that SDT only introduces at most 2\% extra overhead than full testbeds on multi-hop latency and is far more efficient than software simulators (reducing the evaluation time by up to 2899x). SDT is more cost-effective and scalable than existing Topology Projection (TP) solutions. Further experiments show that SDT can support various network research experiments at a low cost on topics including but not limited to topology design, congestion control, and traffic engineering. 

\end{abstract}

%% file: body.tex
\section{Introduction}

As the main bottleneck of Data Centers (DCs), the Data Center Networks (DCNs) have attracted much research attention from both industry and academia~\cite{bergman2018empowering}. There exist some commonly used DCN topologies that are scalable and cost-effective including Fat-Tree~\cite{al2008scalable}, Dragonfly~\cite{kim2008technology}, Torus~\cite{adiga2005blue}, BCube~\cite{guo2009bcube}, HyperBCube~\cite{lin2012hyper}, et al. Research on DCNs, including congestion control mechanisms, routing algorithms, deadlock avoidance functions, et al., should be applied to most of these topologies (or at least some) for better generality (e.g.,~\cite{wu2022detecting, stephens2014practical}). There are also many pieces of state-of-the-art research on optimizing the physical topology to improve the application performance like Distributed Machine Learning (DML)~\cite{wang2019impact}. All of these require a testbed that can support multiple topologies to verify the effects of each mechanism.

\begin{figure}[tbp]
\centerline{\includegraphics[width=9cm]{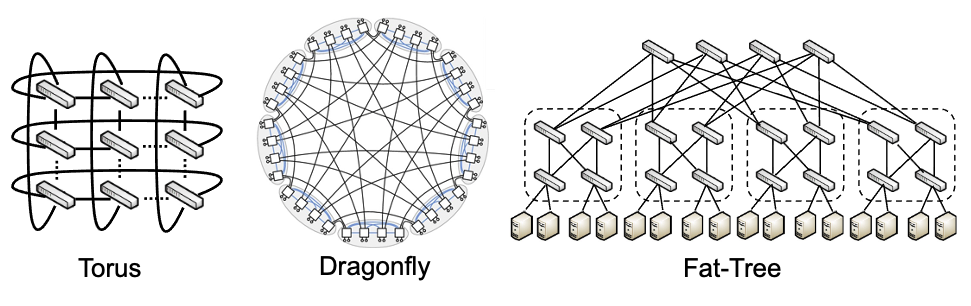}}
\caption{Commonly used topologies}
\label{fig:state-of-art-topologies}
\end{figure}

It is not easy to support multiple topologies at the same time and do reconfiguration among them. First, building a topology such as Fat-Tree can be complex. For example, it needs 20 4-port switches and 48 cables to deploy a standard Fat-Tree topology supporting only 16 nodes (Figure~\ref{fig:state-of-art-topologies}). In addition, it is more complicated to support different topologies and reconfigurations simultaneously. Connections are error-prone and difficult to check when reconfiguring. Although emulators (e.g., Mininet~\cite{web:mininet, lantz2010network}, Open vSwitch~\cite{pfaff2015design}, OpenStack~\cite{web:openstack}) can simulate a variety of topologies, they still have some obvious drawbacks such as long simulation time and insufficient authenticity of results. Therefore, deploying a full testbed for evaluation is crucial and irreplaceable, even if it is hard to make. 

As far as we know, a qualified real-world testbed requires several characteristics, including fast topology reconfiguration, cost-friendly deployment, and convenient maintenance. The challenges in designing such a testbed lie in how to support topology reconfiguration, preferably without manual switching of cables; how to reduce the cost of the test platform, including hardware and labor costs; and even how to support user-defined topologies, rather than being limited to the existing commonly used topologies.

Switch Projection (SP) is a solution to construct topologies for network experiments but needs heavy staffing. The good news is that the Micro Electro Mechanical System (MEMS) optical switches can be used to build reconfigurable network topologies~\cite{wang2010c, farrington2010helios}. Based on its reconfigurable and lossless bi-switching property, it can take the place of SP's manpower. We call the SP with MEMS optical switches the ``Switch Projection-Optical Switch (SP-OS)''. SP-OS can construct user-defined topologies and support real-time reconfiguration without manual operations. However, it still has certain disadvantages, such as high cost and poor expandability. Considering the above characteristics and challenges, we propose a topology-reconfigurable testbed named \emph{Software Defined Topology Testbed (SDT)} without costly optical switches to achieve lower cost and better scalability. 

In short, the contributions of the paper are

\begin{figure}[tbp]
\centerline{\includegraphics[width=8cm]{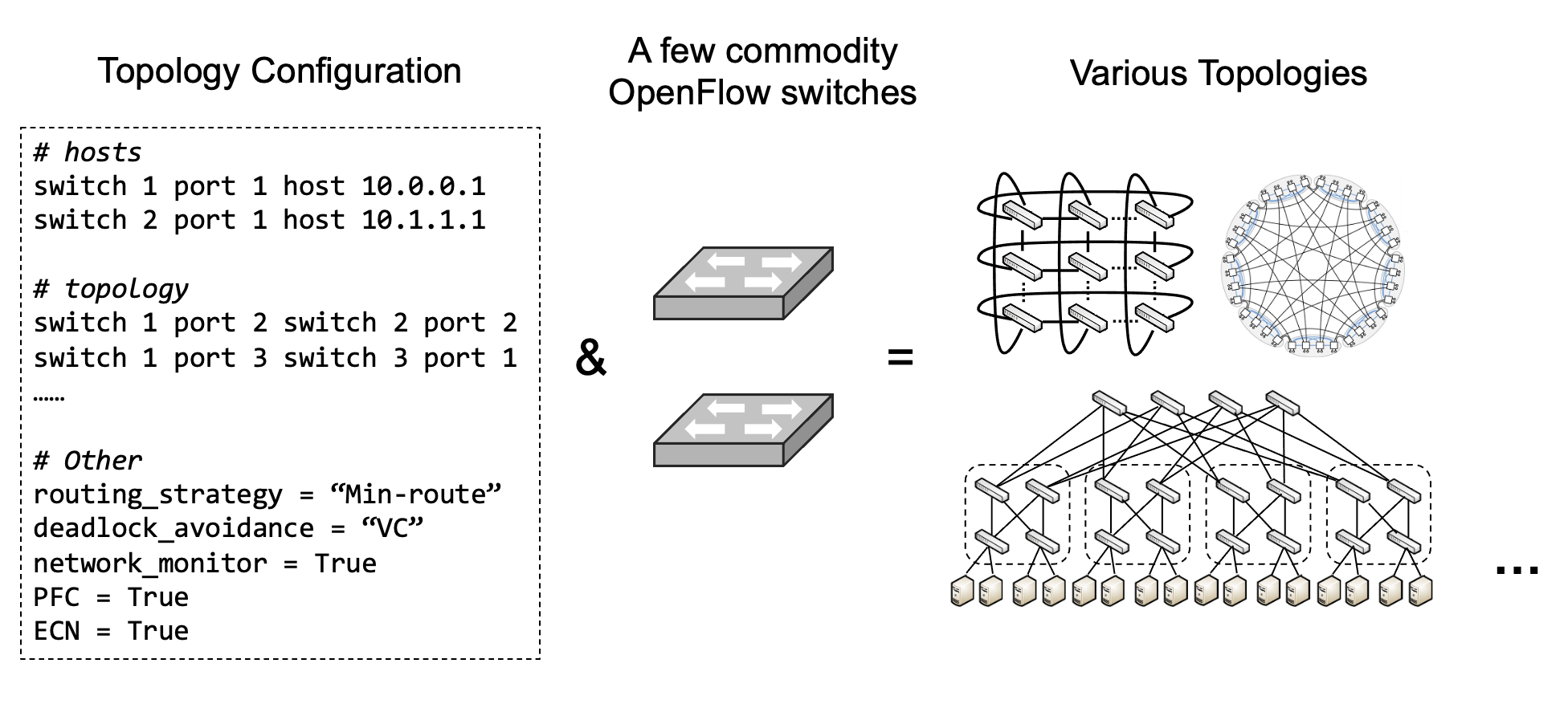}}
\caption{What can SDT do?}
\label{fig:sdt-big-pic}
\end{figure}

\begin{itemize}
  \item We summarize the methodology of Topology Projection (TP) and propose SDT, a testbed solution for building real topologies. SDT uses commodity OpenFlow switches to construct various topologies. Once the connection deployment is completed, the topology (re)configuration can be finished in a short time without manually changing the physical connections or using optical switches (Figure~\ref{fig:sdt-big-pic}). 
  \item We develop an easy-to-use SDT controller supporting user-defined topologies. Users can develop their routing strategy or other new technologies with the SDT controller. The transformation process from logical topology to physical topology is fully automated.
  \item We compare SDT with existing TP methods, and SDT shows better cost-effectiveness and scalability. We use real applications to evaluate 1) the latency and bandwidth differences compared with the full testbed and 2) the Application Completion Time (ACT) and time consumption compared with the simulator. Evaluations show that SDT has only 0.03-2\% deviation on latency compared to the full testbed and reduces the evaluation time by up to 2899x faster than the simulator in a 16-second HPC benchmark for communication efficiency with 32 nodes.
  \item We further implement some prevalent network functions on SDT, including routing strategy, deadlock avoidance, and congestion control. SDT shows substantial flexibility in network evaluations.
\end{itemize}

The rest of the paper is organized as follows. We introduce the related works in \S~\ref{sec:related-works}. We present the motivation and design of SDT in detail in Sections \ref{sec:motivation} and \ref{sec:sdt-design}. A prototype of SDT controller is introduced in \S~\ref{sec:implementation}. The accuracy and efficiency of SDT are evaluated in \S~\ref{sec:evaluation}, with some state-of-the-art network functions implemented. We discuss SDT in \S~\ref{sec:discussion} and conclude the paper in \S~\ref{sec:conclusion}.

\section{Related Works}\label{sec:related-works}

\subsection{Reconfigurable Networks}\label{sec:reconf-network}

To better allocate link bandwidth in response to the non-uniform traffic often present in DCNs, some researchers propose reconfigurable networks, which can dynamically adjust links based on real-time network traffic to better serve hot node pairs (nodes with heavy traffic). These reconfigurable networks are often implemented with optical devices, which can offer lossless bi-switching capabilities. The optical devices used in reconfigurable networks can mainly be categorized into MEMS-based optical switches and other specialized optical devices (e.g., free-space optics and optical devices that forward based on light wavelength). 

\subsubsection{Reconfigurable Networks based on MEMS Optical Switch}

MEMS optical switches use several tiny mirrors on the silicon crystal to forward the light between different fiber interfaces. The tiny mirrors are called microarrays, working as a reconfigurable static crossbar by rotation. 

MEMS optical switches have been put into practical usage very early, and the technology is relatively mature and less error-prone. Therefore, early reconfigurable networks, such as c-Through~\cite{wang2010c} and Helios~\cite{farrington2010helios}, use MEMS optical switches to build reconfigurable networks. However, MEMS optical switches still have drawbacks, such as their relatively large reconfiguration delays (about $100ms$) and high hardware costs.

\subsubsection{Reconfigurable Networks based on Customized Optics}

To achieve faster reconfiguration, researchers have proposed other customized optical devices, such as Free Space Optics used in Firefly~\cite{hamedazimi2014firefly} and ProjecToR~\cite{ghobadi2016projector}, which reflect the laser propagating in the air with mirrors that can do faster angle adjustment to complete the reconfiguration. This kind of network can achieve reconfiguration as fast as $12\mu s$, but it is easily disturbed by the environment, which causes significant optical path shifts and makes the deployment impossible. 

In addition, Sirius~\cite{ballani2020sirius} uses Arrayed Waveguide Grating Router (AWGR) to forward the input light of different wavelengths to the corresponding output ports to complete the reconfiguration. However, this method needs to be used with a highly customized tunable laser that can quickly generate lasers of different wavelengths, which is also less practical.

Besides these, there are some other similar customized-optics-based fast reconfiguration works like~\cite{mellette2017rotornet, mellette2020expanding}.

\subsection{Network Evaluation Tools}

Network researchers have developed and used many network evaluation tools in the past few decades. We roughly divide them into 1) simulator, 2) emulator, and 3) testbed. They have played a significant role in the progress of network technologies, but they also have certain disadvantages. 

\subsubsection{Simulator}

Existing network simulation tools such as NS-2~\cite{web:ns2}, NS-3~\cite{web:ns3}, OPNET~\cite{web:opnet}, OMNET++~\cite{web:omnetpp} and GloMoSim~\cite{web:glomosim} offer efficient and cost-effective ways to evaluate the network performance under different conditions. However, compared with the testbed, they lack both scalability and reality. Simulators may take several days to complete one simulation, and they also suffer from the lack of ability to simulate various random situations that might occur in real networks.

\subsubsection{Emulator}

The primary goal of network emulators such as Mininet~\cite{web:mininet, lantz2010network} with Open vSwitch (OVS)~\cite{pfaff2015design} and Netem~\cite{web:netem} is to create an environment whereby users can flexibly combine the VMs, applications, products, and services to perform a relatively more authentic simulation. However, the performance of emulators is poor in the high bandwidth environment (10Gbps+) or medium-scale topologies (containing 20+ switches) due to the limitation of the system resources. Besides, emulators cannot do everything we want, e.g., Mininet has no official support for Priority-based Flow Control (PFC), even though PFC is already a standard feature.

As a widely used cloud computing infrastructure software, OpenStack~\cite{web:openstack} can be used to build a set of computing nodes with specific topologies using commodity servers and switches. However, the construction of topology on OpenStack is still virtualized by OVS. As a result, the network topology on OpenStack has scalability and reality problems and will be limited by the bandwidth.

\subsubsection{Testbed}

Existing testbed platforms available to researchers include Emulab~\cite{hibler2008large}, CloudLab~\cite{duplyakin2019design} and PlanetLab~\cite{chun2003planetlab}, which have made considerable progress in making testbed as easy to use and control as simulation. Nevertheless, their drawbacks are also obvious. Whether virtualization is used or not, the reconfiguration of the testbed requires heavy manual operations. Several testbeds dedicated to wireless environments are proposed, such as TWIST~\cite{handziski2006twist}, and DRIVE~\cite{pinart2008drive}. These works mainly consider wireless environments, which do not apply to DCN-related experiments.

\section{Motivation and Background}\label{sec:motivation}


This section firstly introduces our motivation for \textit{``Topology Projection (TP)''}. Then, we summarize a straightforward solution named Switch Projection (SP). The SP can support TP easily but can not be reconfigured without manpower. MEMS optical switches can be introduced for topology reconfiguration, which is introduced at the end of this section with the name Switch Projection-Optical Switch (SP-OS).

\subsection{Why Do We Need the SDT?}

By comprehensively considering the pros and cons of three types of existing network evaluation tools (Table~\ref{tab:comp-eva-tools}), we find that they are generally unable to achieve high-performance and low-cost evaluations for various network topologies. Although the simulation is easy to operate and the cost is relatively small, its scalability is limited by the high time cost. As the number of nodes increases and the network traffic grows, the simulation time can be thousands of times longer than the real-world ACT. Testbeds are needed to get better evaluation scalability and efficiency. However, the deployment expenses of testbeds are high and even unacceptable for researchers. 

\begin{table}
\centering
\setlength{\tabcolsep}{0.9mm}{
\begin{tabular}{@{}lllll@{}}
\toprule
 & Simulator & Emulator & Testbed & SDT \\ \midrule
Price & Low & Medium & High & Medium \\
Manpower & Low & Low & High & \textbf{Low} \\
(Re)configuration & Easy & Medium & Hard & \textbf{Easy} \\
Scalability & Low & Medium & High & \textbf{High} \\
Efficiency & Low & Medium & High & \textbf{High} \\ \bottomrule
\end{tabular}}
\caption{Comparison of Network Evaluation Tools for Various Topologies}\label{tab:comp-eva-tools}
\end{table}

Therefore, we want to construct a system that performs almost the same as the full testbed with high efficiency and scalability. The system should support fast reconfiguration among various topologies without changing the physical connections under an acceptable budget. That is why we present SDT. The efficiency of SDT is close to full testbeds without any manual operation during reconfiguration and with lower hardware costs.

\subsection{A Possible Solution: Switch Projection}\label{sec:sp}

Some works (e.g.,~\cite{goyal2022backpressure, jin2017netcache}) use a switch to construct a simple topology for evaluation. We call this method of constructing a topology ``TP''. SDT is also a TP method.

\begin{figure}[htbp]
\centerline{\includegraphics[width=9cm]{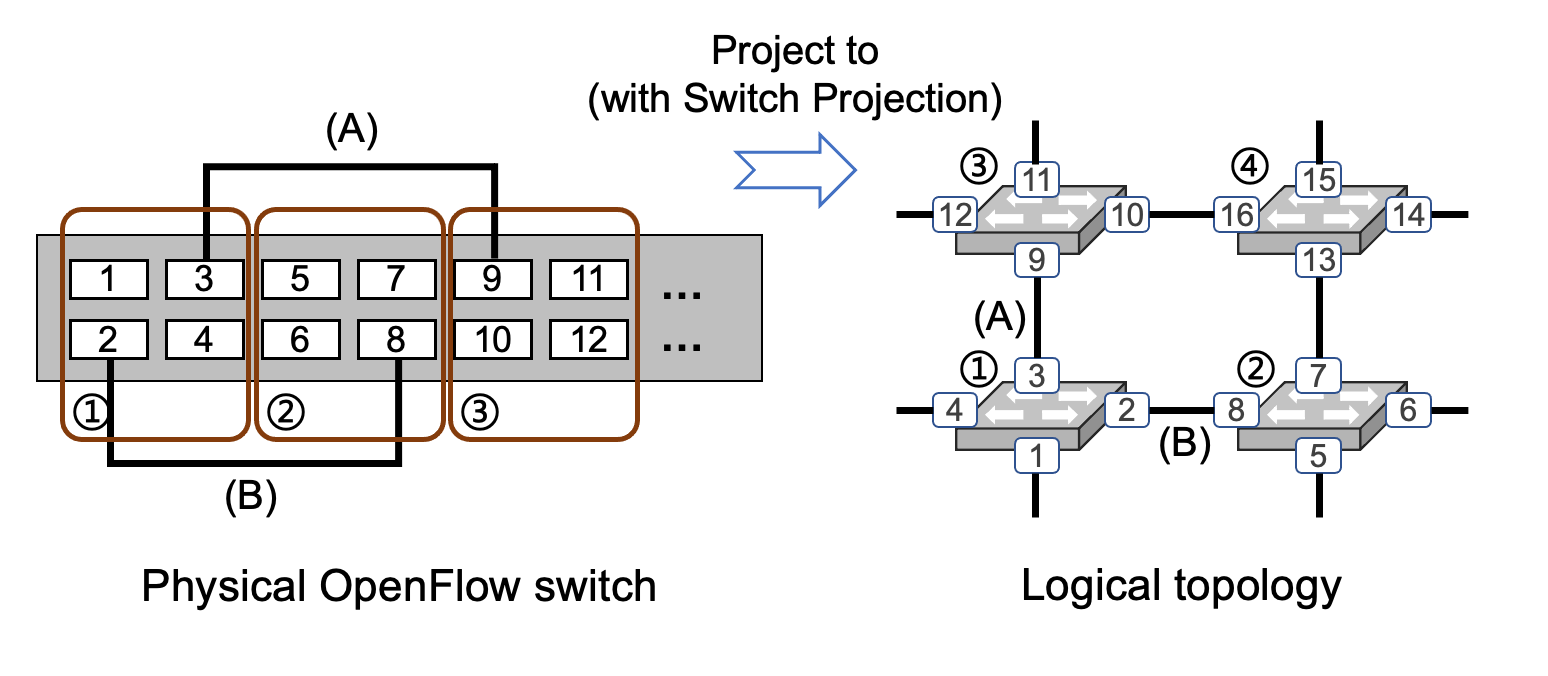}}
\caption{An Example of Switch Projection.}
\label{fig:conv-tp}
\end{figure}

The main idea of traditional TP is to project the topologies by using the logical switch as a meta unit. The right side of Figure~\ref{fig:conv-tp} is the topology we want to construct, which is a part of a 2D-Torus. We call this ``logical topology''. The radix of the switches in this logical topology is 4, i.e., every logical switch has 4 ports. The physical switch can be divided into sub-switches based on the radix. As a result, each sub-switch has 4 ports as well. After that, we can use these sub-switches for the topology projection.

We call this type of TP ``SP'' and conclude its general approach here. The first step of SP is dividing one physical switch into multiple sub-switches. Then we project the sub-switches to the logical switches in the topology, which is why this method is called SP. After the projection, we manually connect these sub-switches' corresponding ports to build the topology. We can use Software-Defined Networking (SDN) functions (e.g., flow tables in the OpenFlow switch) to divide the sub-switches.

Take Figure~\ref{fig:conv-tp} as an example of how SP works. We first divide and project the sub-switches. Ports 1-4 on the physical switch are considered on one sub-switch, so we project them to an arbitrary logical switch e.g., switch \textcircled{1}. Ports in the logical switch \textcircled{1} are numbered based on the projected ports from the physical switch. The operations are the same for other sub-switches.

We then connect the cables between specific sub-switch ports based on the logical topology. For example, in the logical topology, there is a link between ports 3 and 9 (i.e., Link (A)). We connect the corresponding ports on the physical switch. After all the links are made, it is time to deploy the flow table (we use OpenFlow in this paper) to restrict the packet forwarding domain on the physical switch based on the ports' labels. For instance, data packets entering port 1 can only be forwarded to ports 2-4. The restrictions are based on the partition of sub-switches. 

\subsection{Make SP Topology-reconfigurable}

The manual operations required for SP on topology reconfiguration are massive. We have to re-connect the cables manually on every topology reconfiguration, which is error-prone. As the topology size increases, the difficulty of deployment increases correspondingly. Therefore, we introduce MEMS optical switches into SP to reduce labor costs. The new design is called SP-OS.

\begin{figure}[tbp]
\centerline{\includegraphics[width=9cm]{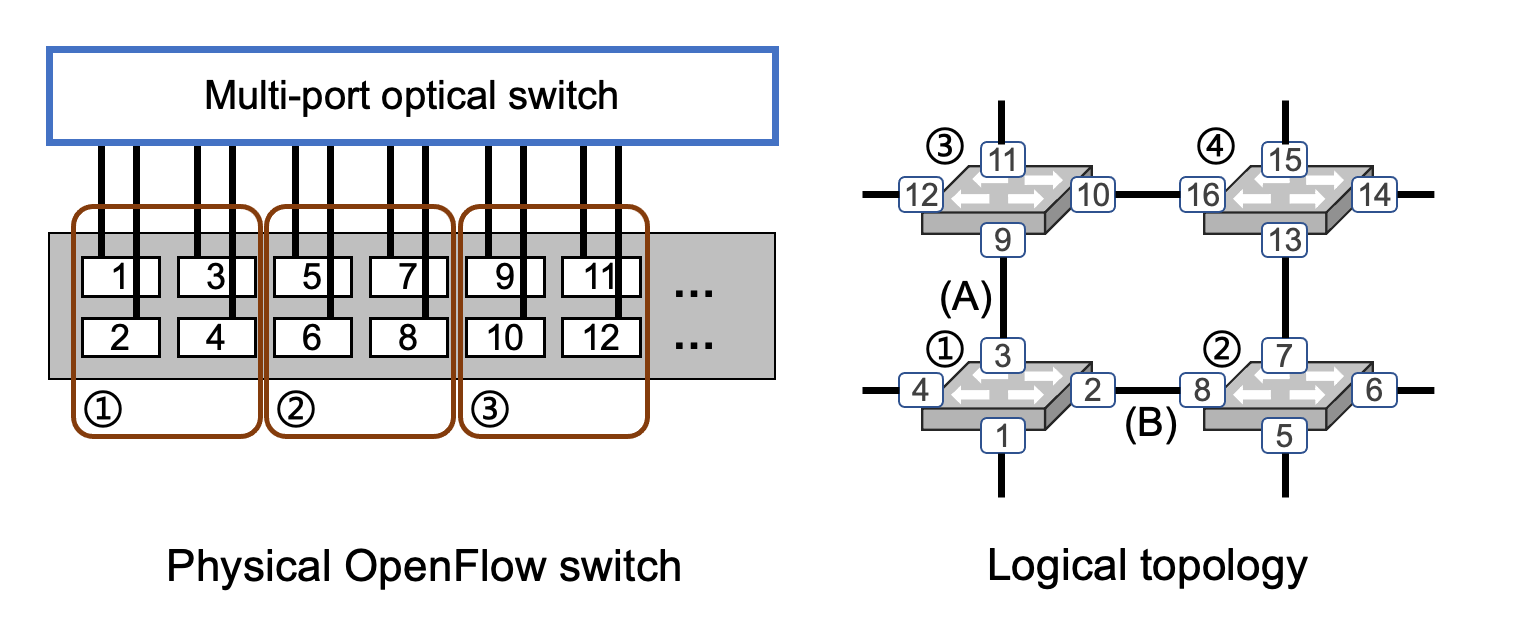}}
\caption{An Example of SP-OS.}
\label{fig:sp-os}
\end{figure}

The optical switch can replace manual operations on the reconfiguration. We connect all the ports on the physical switch to the optical switch (Figure~\ref{fig:sp-os}). When the topology needs to be reconfigured, modifying the configuration of the optical switch based on the labels can replace the manual operations. The advantage of SP-OS is that once the testbed is deployed, all reconfigurations can be done remotely by software control.

The introduction of optical switches leads to increased hardware costs. Optical devices are generally costly. The price of a 320-port MEMS optical switch is more than \$100k, and only 160 LC-LC\footnote{Lucent Connector (LC).} fibers can be connected. As the number of ports on the optical switch increases, the price increases significantly. SDT can work without optical switches, which provides significant savings.

TurboNet~\cite{cao2022turbonet} is another topology-reconfigurable SP method for TP, which replaces manual reconnection with the Tofino switch's loopback ports. However, the use of loopback ports results in a reduction in the available bandwidth of the switches~\cite{de2021flare}. We compare the scalability between TurboNet and SDT in \S~\ref{sec:evaluation}.

\section{The Design of SDT}\label{sec:sdt-design}

In this section, we first introduce the fundamental design of SDT on a single switch. Then, we expand the SDT to multiple switches to support larger topologies. We also address the issue of topology partitioning in multi-switch deployments.

\subsection{SDT on a Single Switch}\label{sec:core_idea_of_sdt}

Although SP-OS can support automated topology reconfiguration, its cost is relatively high due to the introduction of optical switches. Therefore, we design the SDT, which can provide the same functionality as SP-OS but without optical switches. 

\begin{figure}[tbp]
\centerline{\includegraphics[width=9cm]{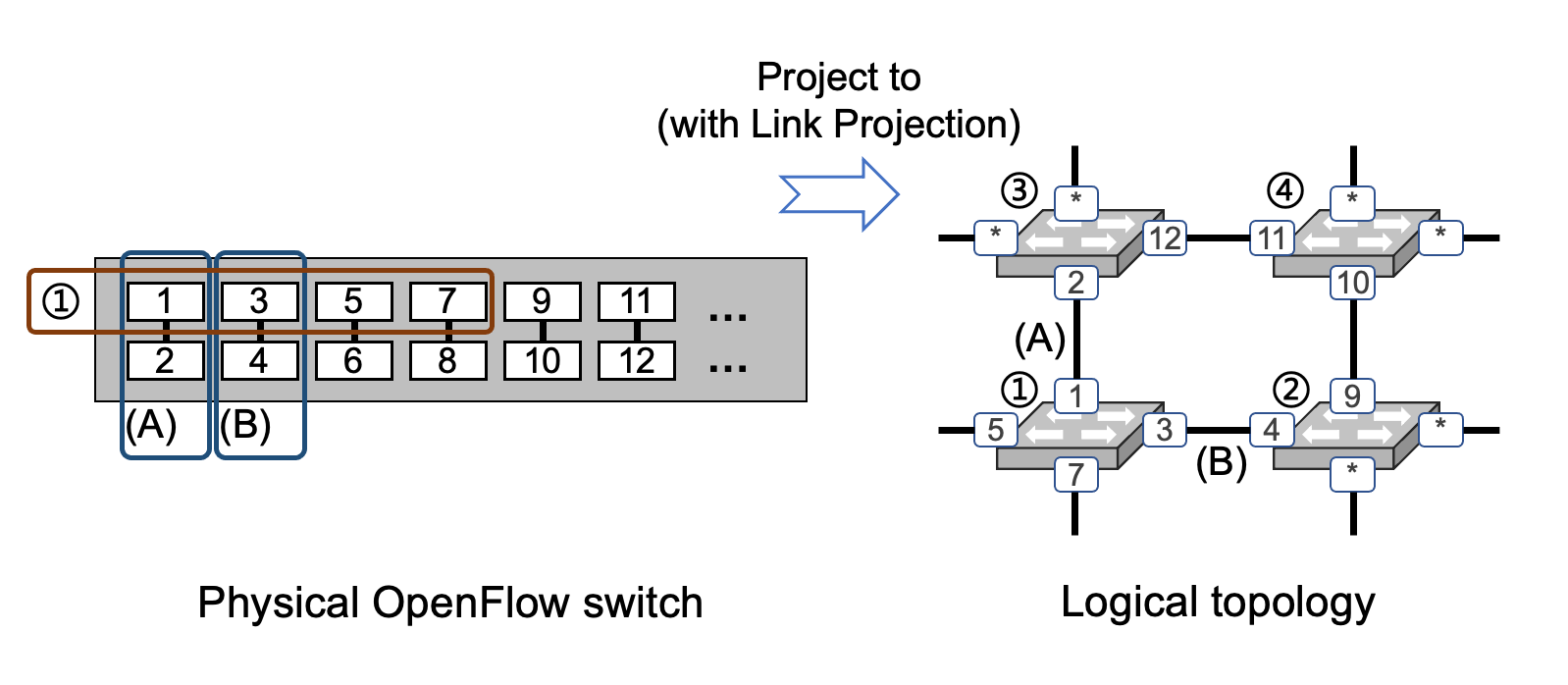}}
\caption{An example of SDT on a single physical switch.}
\label{fig:links-of-sdt-single}
\end{figure}

The main idea of SDT is to use \textbf{Link Projection (LP)} rather than SP to construct the logical topology on a physical switch. SDT first projects physical links\footnote{To construct a physical link, we connect two arbitrary ports on the switch. In the paper, the switch's upper and lower adjacent ports are connected for simplicity.} to logical ones on the topology, and then number the ports on the logical topology based on the projected ports from the physical switch. Taking Figure~\ref{fig:links-of-sdt-single} as an example, the physical links A and B are projected to the logical topology, and then the corresponding ports in the logical topology can be tagged with 1, 2, 3, and 4, respectively.

After the projection, we group ports on the physical switch into different sub-switches based on the relationship of their counterparts in the logical topology. For instance, in Figure~\ref{fig:links-of-sdt-single}, ports 1, 3, 5, and 7 in the topology form a logical switch, so the corresponding ports 1, 3, 5, 7 in the physical switch should be grouped in the same sub-switch. We use OpenFlow flow tables to keep the packets entering this sub-switch only forwarded to their corresponding forwarding domain. The other sub-switches are divided according to these steps as well. 

Please note that no optical switch is needed when the topology is reconfigured in SDT. 

Here we summarize the fundamental differences between SP-OS and SDT.

\begin{itemize}
  \item In SP-OS, sub-switch partitions are determined arbitrarily (the only constraint is that the radix of sub-switches should match the radix of logical switches in the topology). MEMS optical switches are used to (re)connect links between these sub-switches based on the topology's logical switches (projected by \textbf{SP}). 
  \item In SDT, physical links on the physical switch will remain fixed once constructed (which can be arbitrary). The sub-switches are (re)partitioned based on the result of \textbf{LP}. Rules in the flow tables of the OpenFlow switch can be used to realize the sub-switch partition, and no optical switch is needed during a topology reconfiguration.
\end{itemize}

The size of the logical topology supported by SDT is limited by the number of ports on the physical switch. A topology can be appropriately built if the total number of ports in the topology is less than or equal to the number of ports on the physical switch (excluding the ports connected to the end hosts). This constraint applies to all TP methods.

\subsection{SDT on Multiple Switches}\label{sec:multi-sdt}

When one switch is insufficient to project the entire logical topology, multiple switches are needed to use. In SP-OS, it is not difficult to expand the supported logical topology by adding more switches and optical devices. The expansion of SDT is also relatively simple but requires additional discussions below. 

On the construction of the multi-switch scenario, it needs to cut the logical topology into various sub-topologies, and each sub-topology is maintained independently by one physical switch. 

\begin{figure}[tbp]
\centerline{\includegraphics[width=9cm]{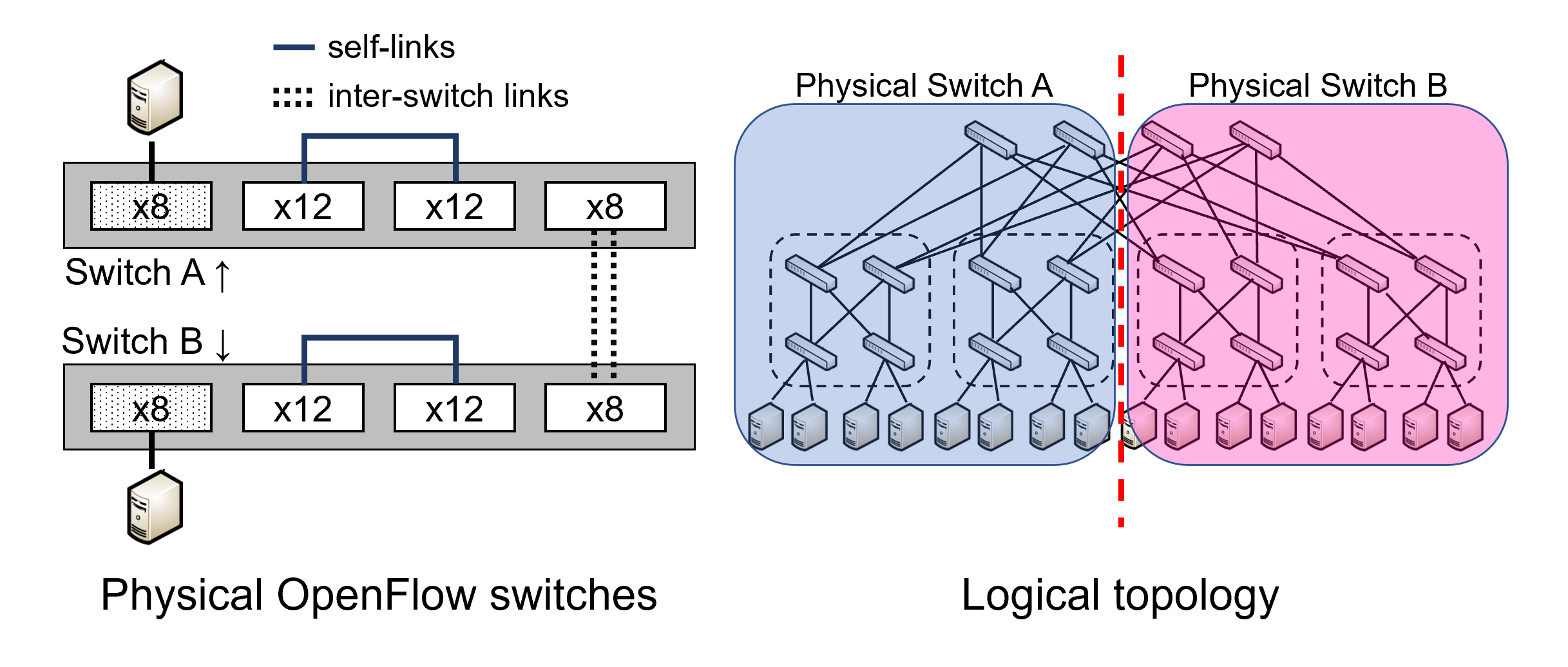}}
\caption{An example of SDT on multiple switches.}
\label{fig:multi-switch-sdt}
\end{figure}

There are two different types of links in multi-switch SDT. We call the links between the upper and lower adjacent ports of one switch \textit{self-links}. For those links across the sub-topologies, we project them from the links across physical switches and call them \textit{inter-switch links}. For instance, the topology has been cut into two sub-topologies on the right side of Figure~\ref{fig:multi-switch-sdt}. The links inside each sub-topology are self-links, and the links between the two sub-topologies are inter-switch links. 

There is a requirement for the number of inter-switch links. Taking Figure~\ref{fig:multi-switch-sdt} as an example, the scale of the logical topology is larger than the previous one. As a result, one 64-port switch cannot build this topology, but two can make it. To build the topology, we divide the topology into two sub-topologies. How to divide the topologies is discussed in Sec. \ref{sec:general-partition-method}.

Here we use the formula to represent the inter-switch links. Define topology (graph) $G(E, V)$ as the logical topology we hope to build, and the sub-topologies are $G_A(E_A, V_A)$ and $G_B(E_B, V_B)$. $E_{nA}$ represents the links to nodes on the physical switch A, $E_{sA}$ represents the self-links on the physical switch A, and $E_{aAB}$ represents the inter-switch links between the physical switches A and B. In the logical topology, there is a relationship: $E = E_n + E_s$. For sub-topologies after being divided, they have

\begin{equation}
    \begin{cases}
    E_A &= E_{nA} + E_{sA} \\
    E_B &= E_{nB} + E_{sB} \\
    V &= V_A + V_B \\
    \end{cases}
\end{equation}

For inter-switch links, the following equation exists.

\begin{equation}
E_{aAB} = E_{aBA} = E - E_A - E_B
\label{equ:inter-switch-links}
\end{equation}

We can now determine the number of inter-switch links for the logical topology by Eq. \ref{equ:inter-switch-links}. For the case in Figure~\ref{fig:multi-switch-sdt}, there are 8 inter-switch links between the two sub-topologies, which means at least 8 inter-switch links are required to construct this topology. 

\begin{figure}[tbp]
\centerline{\includegraphics[width=9cm]{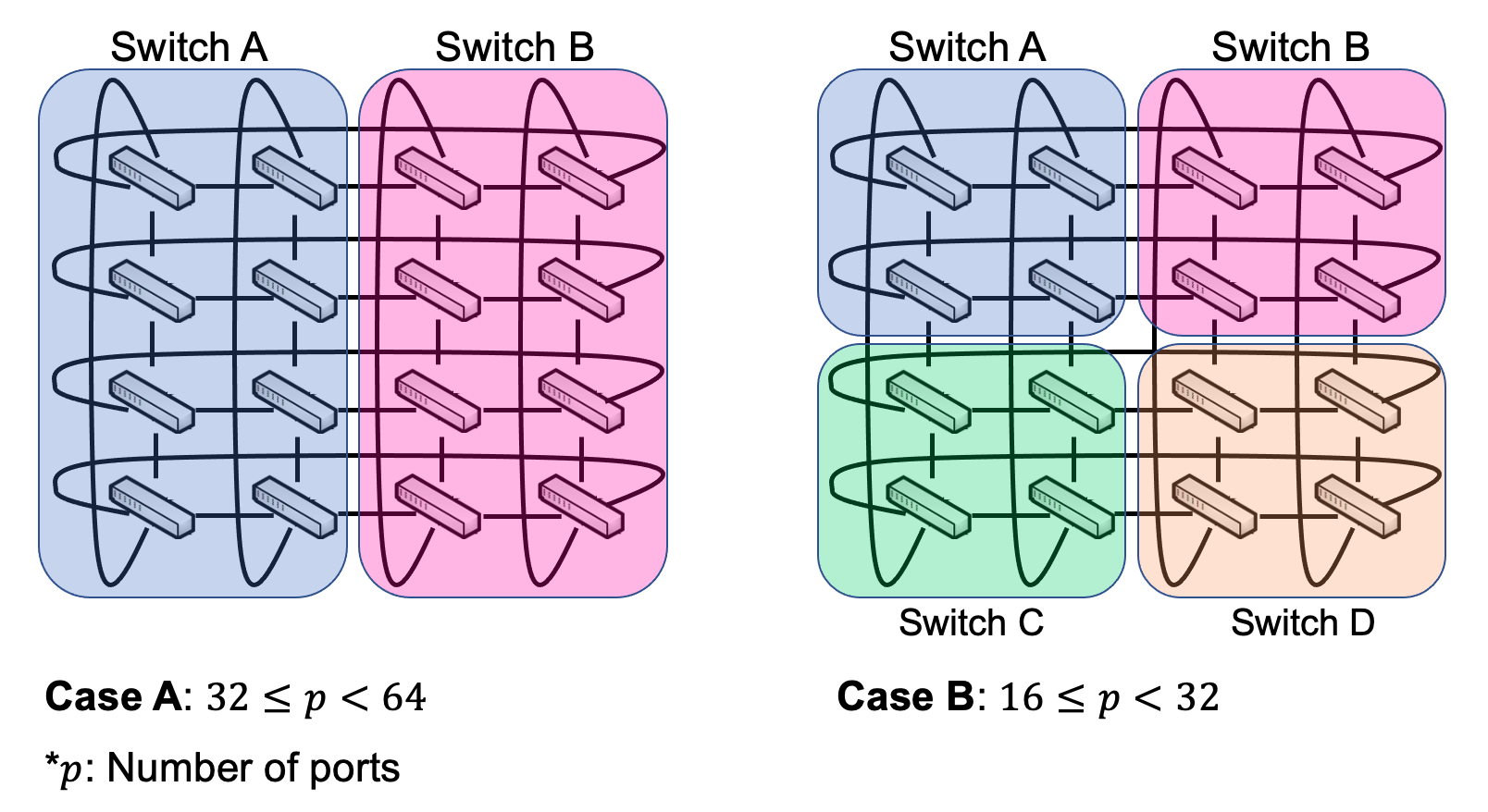}}
\caption{Cases of SDT on multiple switches.}
\label{fig:cases-of-multi-sdt}
\end{figure}

The reservation of inter-switch links is flexible, but it must fulfill the requirements of the desired topologies and the specifications of physical switches. Taking Figure~\ref{fig:cases-of-multi-sdt} as an example, we aim to construct a 4x4 2D-Torus topology (the connections to nodes are omitted for simplicity). When the number of ports on physical switches is greater than 64, only 1 switch is necessary. When the number of ports exceeds 32 but is less than 64, 2 switches are required to build the topology, as shown on the left side of Figure~\ref{fig:cases-of-multi-sdt}. Each switch is assigned 12 self-links and 8 inter-switch links in this scenario. When the number of ports is less than 32 but greater than 16, we can build it with 4 switches. Attention must be paid to determining the switches at both ends of the inter-switch links according to the partition results.

It is worth noting that even if the partitioning methods are different, the results of TP are almost the same. Nevertheless, a proper cutting method enables the testbed to support more topologies without manual modifications. In the implementation, if it needs to perform experiments on multiple topologies, we generally divide the topologies in advance based on the specifications of switches (port number, limitation of flow table et al.) to obtain a proper number of inter-switch links between different switch pairs, i.e., to keep the number of inter-switch links between multiple different switch pairs about the same. The reserved inter-switch links usually come from the maximum inter-switch links among all topologies.

\subsection{Topology Partition for SDT on Multiple Switches} \label{sec:general-partition-method}

The partition of the logical topology needs to be discussed. We define the function ``Cut(G(E, V), params...)'' for dividing the topology. The input of the function is the logical topology G(E, V), switch parameters, and the number of switches. The output is the partitioning method that satisfies the requirements of all the topologies we aim to build and the number of links of each type to be assigned. The problem is represented with switches and nodes as vertices and logical links as edges. The logical topology can be described as an undirected graph. To achieve the partitioning, we apply a graph partitioning algorithm that splits the graph into sub-graphs. 

The partition of the graph needs to meet certain requirements. The first is that the number of inter-switch links should be small, for the inter-switch links are relatively more complicated than self-links. With this requirement, one initial idea is to use the ``Min-cut'' partitioning algorithm to divide the topology. The target is to minimize the $CutEdges(E_A, E_B) = \sum_{u\in V_A, v\in V_B}{w(u, v)}$. Notes that $w(u, v)=1$.

\begin{figure}[tbp]
\centerline{\includegraphics[width=8cm]{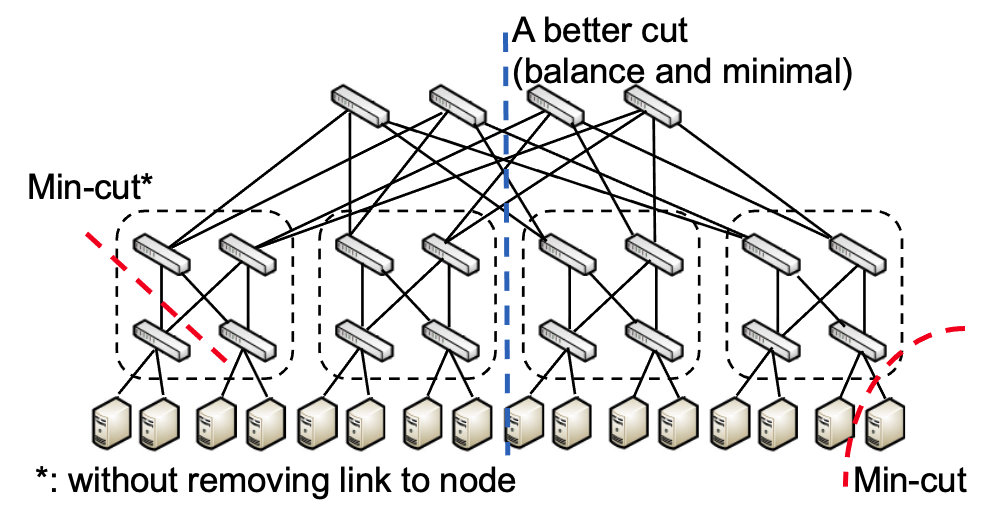}}
\caption{Differences between partitioning methods.}
\label{fig:diff-of-cuts}
\end{figure}

Besides this, we also want to keep the number of used links (or ports) per physical switch as balanced as possible. It is beneficial to balance the number of ports and links of each physical switch in terms of resource usage and complexity of ports to nodes. However, Min-cut partitioning can not work well under this condition. Figure~\ref{fig:diff-of-cuts} shows the differences between these partitioning methods. Another graph partitioning algorithm is needed, whose target is to minimize $\alpha \times Cut(E_A, E_B) + \beta \times (\frac{1}{\sum_{E_A}^i{1}} + \frac{1}{\sum_{E_B}^i{1}})$. 

To summarize the requirements for the SDT partitioning algorithm, the graph partitioning algorithm should 1) minimize the number of edges between sub-graphs and 2) balance the number of edges within each sub-graph. Meeting these requirements is a proven NP-hard problem, and algorithms such as RatioCut~\cite{ratiocut-hagen1992new} or minimize normalized cut (NCut)~\cite{ncut-shi2000normalized} can be used to solve it. In practice, we use the widely-used METIS library~\cite{karypis1998fast} with these constraints to perform the partitioning of the topology, and the results are usually satisfactory. When multiple topologies need to be evaluated in real-world experiments, we perform graph partitioning for all topologies and then select the maximum number of inter-switch links as the reference for deployment on the physical topology.

\section{Implementation Details: SDT Controller}\label{sec:implementation}

\begin{figure}[tbp]
\centerline{\includegraphics[width=9cm]{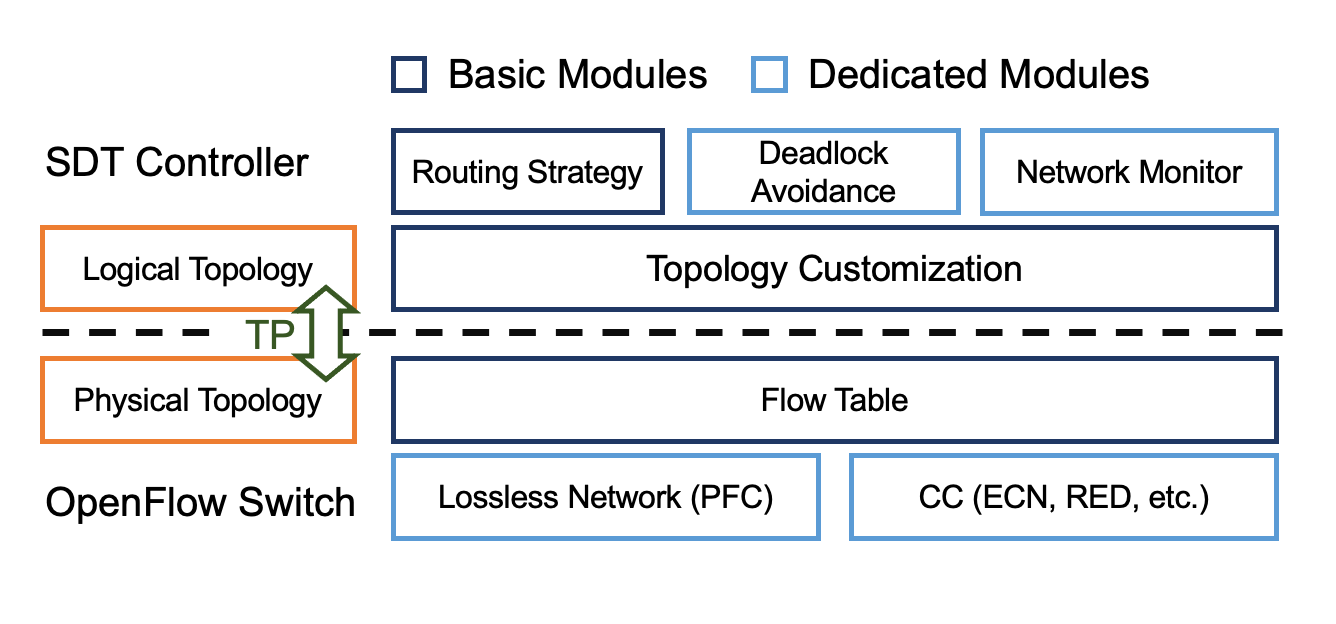}}
\caption{Architecture of SDT controller. }
\label{fig:arch-of-sdt-controller}
\end{figure}

We implement the SDT controller based on the library Ryu~\cite{web:ryu} under version 4.34 and the API in commodity OpenFlow switches. As shown in Figure~\ref{fig:arch-of-sdt-controller}, the SDT controller consists of 4 modules. \textit{Topology Customization} and \textit{Routing Strategy} are two basic modules of the controller. The remaining two modules, i.e., \textit{Deadlock Avoidance} and \textit{Network Monitor}, are dedicated modules for DCNs. SDT controller supports fast (re)configuration of network topology and other modules by running a simple configuration file as shown in Figure~\ref{fig:sdt-big-pic}. 

\subsubsection{Topology Customization}\label{sec:custom-topology}

This module is essential for performing TP, consisting of 1) the checking function and 2) the deployment function. In the checking function, all user-defined topologies will be used as input to the module, along with how the testbed is connected (e.g., distribution of nodes and two types of links). The module first checks if these topologies meet the deployment conditions as addressed in \S~\ref{sec:general-partition-method}. If not, the module will inform the user of the necessary link modification. Then, the checked user-defined topology is used as the input for the deployment function. The controller will maintain the logical topology as an undirected graph and run the TP process automatically in this function. 

\subsubsection{Routing Strategy}\label{sec:route_strategy} 

This module contains various routing strategies for different topologies. We implement several routing algorithms as shown in Table~\ref{tab:routing-strategy}. Most of the user-defined routing strategies can be implemented by the SDT controller as a specific set of flow tables. For instance, when a new flow comes, the SDT controller calculates the paths on the logical topology according to the strategies and then delivers the corresponding flow tables to the proper OpenFlow switches to perform a specific routing for the flow. 

\subsubsection{Deadlock Avoidance and Network Monitor} 
These two modules are dedicated modules for DCNs. The former works in the lossless network, like RDMA over Converged Ethernet (RoCE), along with \textit{Routing Strategy} module to avoid the deadlock. The latter is mainly used for network telemetry. For example, the SDT controller periodically collects statistics data in each port of OpenFlow switches through provided API. The collected data can be further used to calculate the load of each logical switch in the case of adaptive routing.

We use the SDT controller to implement some prevalent network functions to evaluate SDT's capability. For details, please refer to \S~\ref{sec:real_network_functions}.

\section{Evaluation}\label{sec:evaluation}

In this section, we conduct several experiments to answer the questions, including:

\begin{enumerate}
    \item Will SDT introduce additional overhead (e.g., latency) compared to a full testbed? (\S~\ref{sec:evaluation_accuracy})
    \item How many types of topologies can SDT project? (\S~\ref{sec:scalability})
    \item How cost-effective and scalable is SDT compared to previous TP methods? (\S~\ref{sec:scalability})
    \item How much speed-up can SDT bring to network experiments? (\S~\ref{sec:act_comparation})
    \item Can existing network functions be applied to SDT? (\S~\ref{sec:real_network_functions})
\end{enumerate}

It is worth mentioning that all topology reconfigurations of SDT in this section are done remotely without any manual rewiring.

\subsection{Experiment Setup}

\subsubsection{SDT Cluster Setup}

We use 3 H3C S6861-54QF OpenFlow switches (with 64 10Gbps SFP+ ports and 6 40Gbps QSFP+ ports, which can be split into 4 10Gbps SFP+ ports) for SDT. We use 16 HPE DL360 Gen9 servers with E5-2695v4 (18 cores and 36 threads) as host servers and virtualize them to 32 computing nodes (i.e., virtual machines). Each host server has one Mellanox ConnectX-4 10GbE dual-port NIC. Each computing node is allocated with 32GB RAM and 8 CPU cores. Moreover, each computing node is bound with a physical NIC port through SR-IOV to ensure that the virtualization will not become the performance bottleneck. All the network devices support the Priority Flow Control (PFC) for lossless ethernet.


\subsubsection{Baselines}

We use a full testbed to compare the accuracy of SDT in terms of latency and bandwidth. We compare the Application Completion Time (ACT) of SDT with a self-designed simulator running different HPC applications under different topologies. We also evaluate the cost-effectiveness and scalability compared to SP, SP-OS, and TurboNet~\cite{cao2022turbonet}.

The network simulator we use is based on two popular simulators BookSim~\cite{jiang2013detailed} and SST/Macro~\cite{web:sst}. The simulator supports a range of features needed by the evaluations (including PFC, cut-through, trace replaying, et al.) and is event-driven for efficiency. To run the same application as the nodes on SDT, the simulator uses the traces collected from running an HPC application on real computing nodes to ensure the simulator's authenticity. We only compare the SDT to the TurboNet with Port Mapper (PM) because the number of queues on each port in the topology projected by Queue Mapper (QM) is inadequate for experiments inside the DCs.

\subsection{TP Accuracy of SDT}\label{sec:evaluation_accuracy}

\begin{figure}[htbp]
\centerline{\includegraphics[width=7cm]{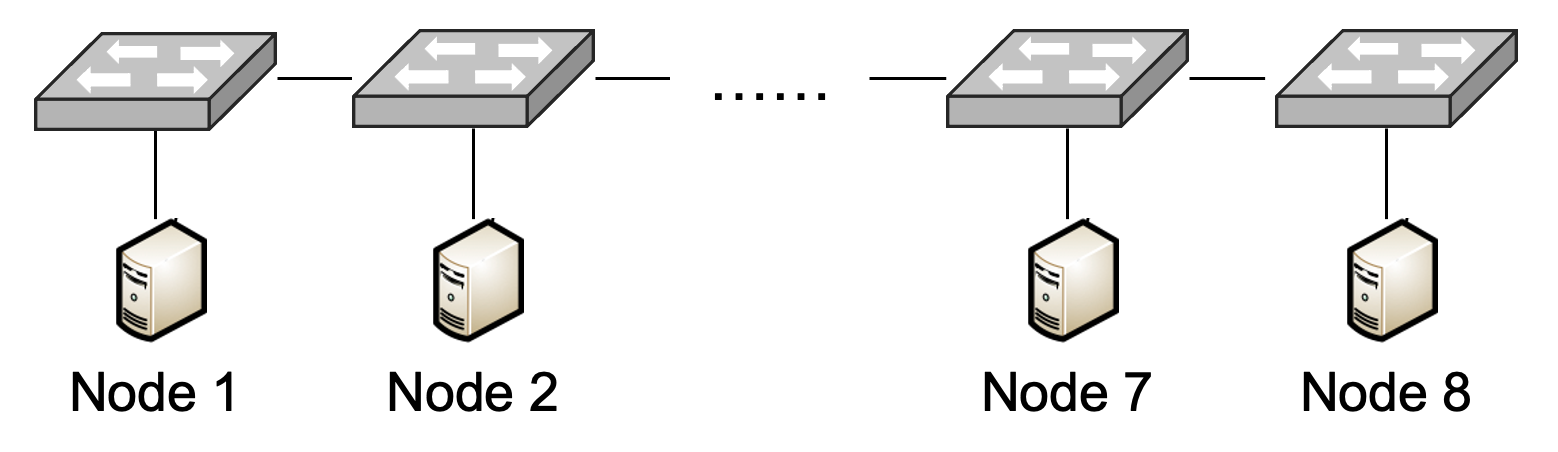}}
\caption{Topology for latency and bandwidth tests}
\label{fig:topo-for-latency-and-bandwidth}
\end{figure}

\subsubsection{Latency} 

We construct a multi-hop topology for latency and bandwidth tests as shown in Figure~\ref{fig:topo-for-latency-and-bandwidth}. The topology consists of 8 switches and computing nodes. There is one node connected to each switch. The switches and nodes are inter-connected with 10Gbps links. We build this topology on SDT and a full testbed and compare the latency between Node 1 to Node 8 by using the Pingpong application in Intel MPI Benchmark (IMB)~\cite{web:imb}. The application is running on the RoCEv2 network with ECN-disabled. 

\begin{figure*}
\centering

\begin{minipage}[t]{5cm}
    \centering
    \includegraphics[width=4.5cm]{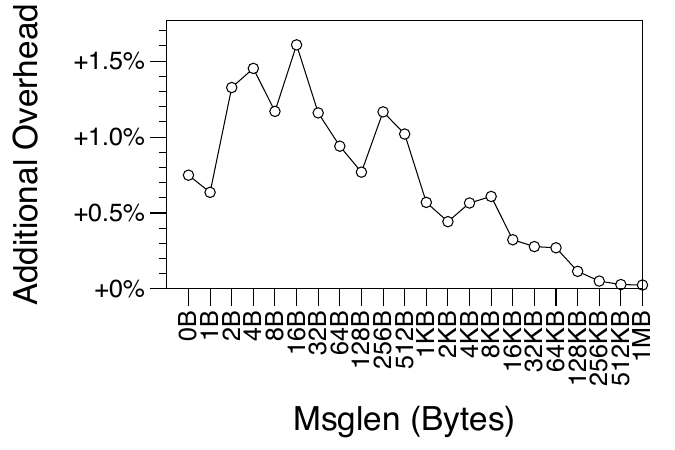}
\caption{Additional overhead by SDT on 8-hop latency.}
\label{fig:latency-test-result}
\end{minipage}%
\begin{minipage}[t]{13.5cm}
    \centering
    \includegraphics[width=13cm]{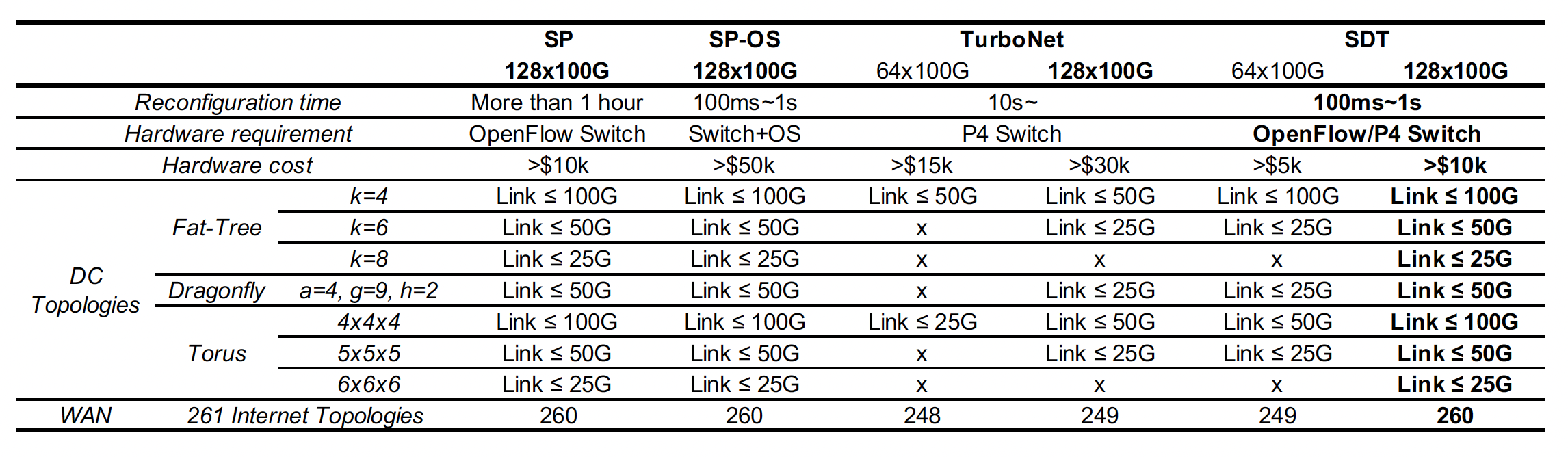}
    \captionof{table}{Comparison between SDT and other TP methods.}
    \label{tab:topology_comparison}
\end{minipage}%
\vspace{-0.5cm}
\end{figure*}

\begin{figure*}[htbp] 
\centering  

\includegraphics[width=18cm]{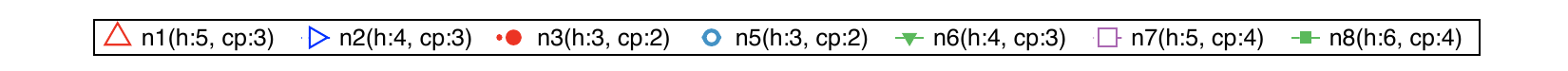}
\quad
\subfigure[SDT (PFC on)]{
\includegraphics[width=4cm]{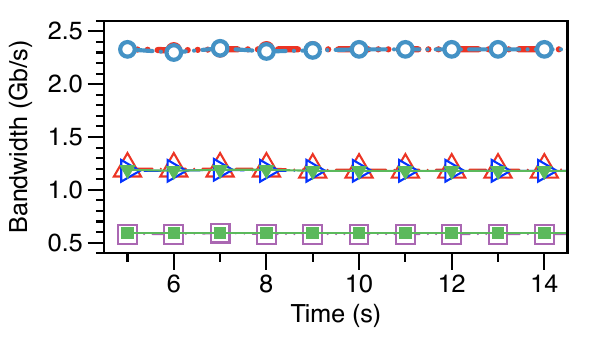}
}
\subfigure[Full testbed (PFC on)]{
\includegraphics[width=4cm]{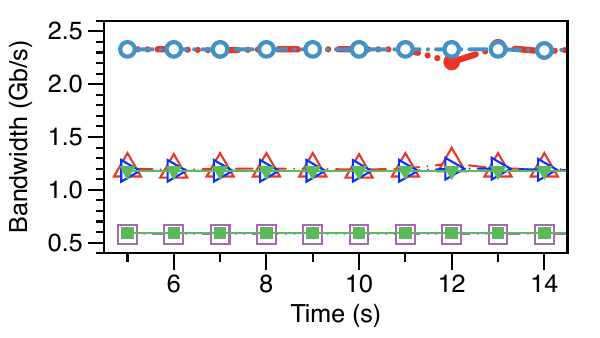}
}
\subfigure[SDT (PFC off)]{
\includegraphics[width=4cm]{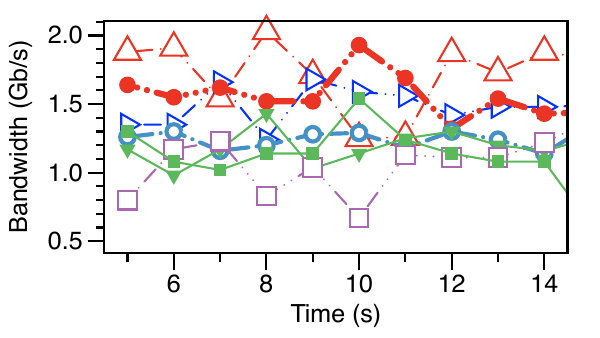}
}
\subfigure[Full testbed (PFC off)]{
\includegraphics[width=4cm]{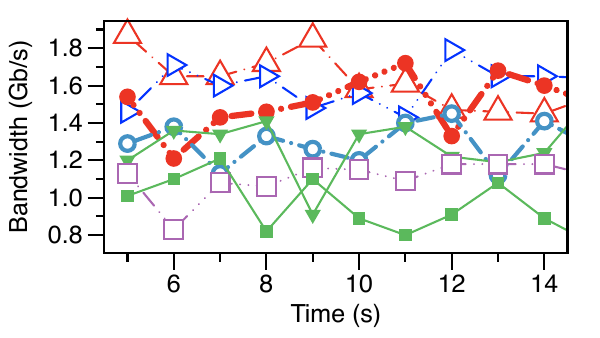}
}

\caption{Bandwidth test results. [Legend: h denotes hop(s), cp denotes congestion point(s). h:3, cp:2 means that this node has 3 hops away from node 4 (target node), and there are 2 congestion points on the way.]}
\label{fig:bandwidth_test_results}

\end{figure*}

We perform the latency test 10k times on incremental message lengths (param -msglen) and collect the latencies. Define the average latency of the full testbed as $l_r$, and the latency of SDT is $l_s$. The overhead is calculated by $\frac{l_s - l_r}{l_r}$. Figure~\ref{fig:latency-test-result} shows that the SDT would bring an acceptable overhead to the RTT. It is worth noting that the latency is quite small in the RoCEv2 network, which means introducing any tiny delay can lead to large deviations in results. For example, the 10-hop latency of the lengths below 256 bytes is under $10\mu s$. Although the latencies on RoCEv2 are sensitive to the hardware conditions, the overheads brought by SDT are below 1.6\%, which can be ignored. With the increment of message lengths, the overhead brought by SDT is getting smaller. 

\subsubsection{Bandwidth} We use iperf3 to construct an incast scenario for bandwidth test: all other nodes send 10Gbps TCP traffics to node 4. We compare the bandwidth on loss and lossless networks (with PFC off/on, respectively).

The results (refer to Figure~\ref{fig:bandwidth_test_results}) demonstrate that with PFC enabled, the bandwidth allocation for each iperf3 flow aligns with the full testbed. For instance, nodes 3 and 5, which have 2 congestion points on their path to node 4, have comparable bandwidth when controlled by PFC in both the SDT and full testbed. Their bandwidth allocation is significantly distinct from that of other nodes with different hop counts. In the network without PFC, the bandwidth distribution between SDT and the full testbed has a nearly identical trend. Nodes that can allocate relatively high bandwidth (which may be influenced by RTT and other factors) behave similarly in both the actual topology and SDT. The trends are nearly alike for nodes with lower bandwidth. The only differences may be due to the additional overhead introduced by SDT, leading to slight differences in RTT and therefore different window growth rates. 

To summarize, the way SDT builds the topology does introduce a bit of additional overhead, resulting in a deviation of 1.6\% or less to the latencies compared to the full testbed in our environment. Our initial speculation is that these additional latency overheads are because TP increases the load of the switch's crossbar, which causes a slight bias compared to the real environment. These deviations are reasonable and have a negligible impact on the bandwidths.

During the evaluation, we also evaluate the hardware isolation using the Wireshark network sniffer on the client side. We deploy two unconnected topologies in one SDT testbed and conduct the same Pingpong experiment separately. The evaluation results show that the client's port does not receive packets from nodes that are not connected in one topology.


\subsection{Scalability, Convenience, and Cost of SDT}\label{sec:scalability}

We use simulations to compare the scalabilities, conveniences, and costs between SDT and other TP methods (SP, SP-OS, and TurboNet~\cite{cao2022turbonet}) on the projection of multiple topologies, including the widely-used topologies in DCs (Fat-Tree, Dragonfly, and Torus) and 261 WAN topologies (comes from the Internet Topology Zoo~\cite{knight2011internet}). The metric of reconfiguration times is calculated by the total time spent from the time the configuration is placed until the network is available. The hardware costs are extrapolated from the current market price of the hardware. 

Table~\ref{tab:topology_comparison} presents the results of the evaluations and shows that SDT can project more topologies than TurboNet at the same hardware cost, making it more scalable and cost-efficient than SP and SP-OS. SP requires manual reconnection, making reconfiguration time-consuming and prone to errors, especially for large topologies. SP-OS incorporates optical switches (OS) to facilitate reconfiguration but suffers from expensive hardware costs. TurboNet employs the loopback port of P4 switches for reconfiguration, resulting in halved bandwidth on the ports and reduced scalability compared to SDT. Also, recompiling the P4 program is time-consuming. SDT is the best option among these solutions due to its excellent scalability and cost-effectiveness.

\subsection{Comparison between SDT, Simulator, and Full Testbed}\label{sec:act_comparation}



\begin{table}
\centering
\setlength{\tabcolsep}{0.5mm}{
\begin{tabular}{@{}lll@{}}
\toprule
Topology    & Routing Strategy                                  & Deadlock Avoidance                                   \\ \midrule
Fat-Tree    & Depth-First Search (DFS)                                & No need                                              \\
Dragonfly   & Minimal routing                   & Changing VC~\cite{dally1993deadlock, kim2008technology} \\
2D-Mesh     & X-Y routing~\cite{wu2003fault}    & By routing                                           \\
3D-Mesh     & X-Y-Z routing~\cite{ahmed2012xyz} & By routing                                           \\
2D/3D-Torus & Clue~\cite{xiang2011efficient}    & By routing and changing VC                           \\ \bottomrule
\end{tabular}}
\caption{Routing strategies and deadlock avoidance schemes we implemented for different topologies.}\label{tab:routing-strategy}
\vspace{-0.5cm}
\end{table}

\begin{figure*}
\centering

\begin{minipage}[t]{13cm}
\centerline{\includegraphics[width=13cm]{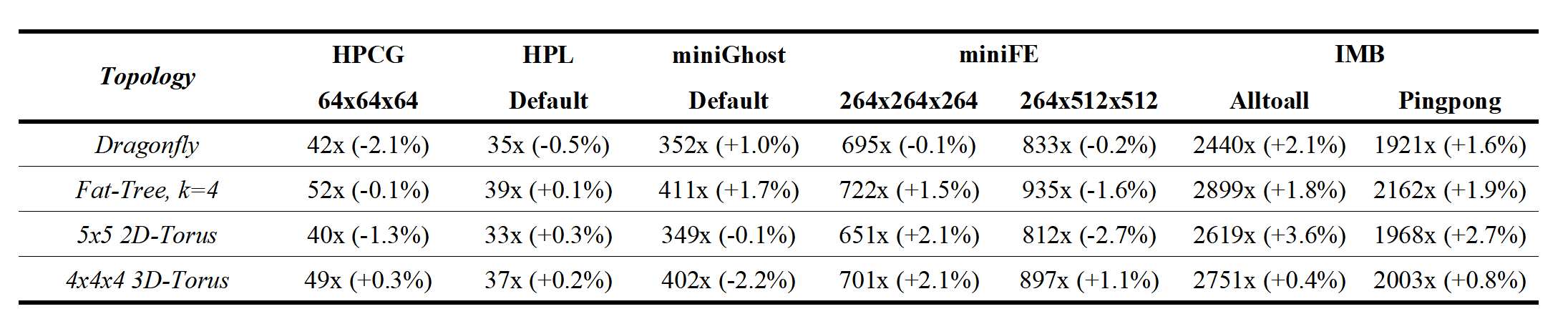}}
\captionof{table}{Real application ACTs on SDT compared to simulator.}
\label{tab:simu-comparison}
\end{minipage}%
\begin{minipage}[t]{5cm}
\centerline{\includegraphics[width=4.5cm]{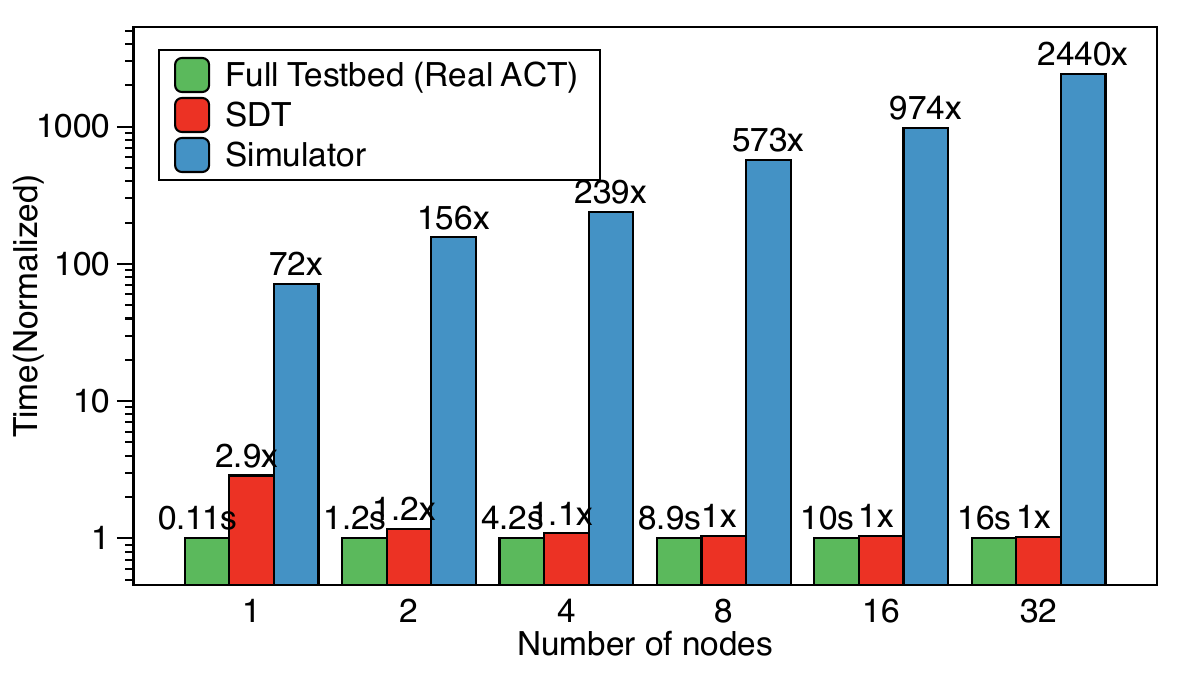}}
\caption{Evaluation times between full testbed, simulator, and SDT.}
\label{fig:time-cost-comparison}
\end{minipage}%
\vspace{-0.5cm}
\end{figure*}

We run a batch of HPC applications and benchmarks, including HPCG, HPL, miniGhost, miniFE, and IMB, to verify the ACT differences among SDT, the simulator, and the full testbed. 
The HPC applications can verify the universality of SDT in the network experiments, while the IMB Alltoall is a pure traffic benchmark without any computation, ideal for verifying the impact on network performances brought by SDT's overhead. We run the applications on specific topologies and construct the topologies on both SDT and simulator. All parameters remain the same for the simulator and SDT, including PFC thresholds, congestion control, DCQCN enabled, cut-through enabled, et al. For details on network functions like deadlock avoidance, please refer to \S~\ref{sec:real_network_functions}.

We select the topologies 1) Dragonfly with a=4, g=9~\cite{kim2008technology}, and h=2, 2) Fat-Tree with k=4~\cite{al2008scalable}, 3) 5x5 2D-Torus, and 4) 4x4x4 3D-Torus~\cite{adiga2005blue} for evaluation. For the topologies with the number of nodes greater than 32, we randomly select the nodes but keep the same among all the evaluations.

Table~\ref{tab:simu-comparison} shows the difference in real-application evaluation between the SDT and simulator. Ax (B\%) in the table represents the evaluation time of the SDT is A times faster than the simulator with a difference of ACT in B\%. The result shows that the ACT collected in SDT is almost identical to the simulator, with a maximum deviation of 3\%. However, the time consumption of SDT is greatly reduced compared to the simulator, especially in applications with heavy traffic.

Further evaluations are conducted to assess the performance improvement brought by SDT as the number of nodes increases. Figure~\ref{fig:time-cost-comparison} compares the time consumption of full testbed (real ACT), simulator, and SDT in evaluating IMB Alltoall benchmark on a Dragonfly topology (a=4, g=9, h=2) with 1, 2, 4, 8, 16, and 32 randomly selected nodes. Note that SDT's time consumption includes the deployment time of the topology. Results show that when the ACT is short, the topology deployment time may result in overhead in the evaluation time consumption, but it is still faster than the simulator. It's worth mentioning that the simulation time may be affected by the performance of the machine running the simulation, but this does not resolve the issue that the simulation is much slower than a real experiment on SDT.

To summarize, SDT can well construct actual network topologies. The experiments performed on SDT show almost the same ACT as the real environments and the simulations, while SDT has much lower costs than the full testbed and is much faster than the simulator. There are good reasons that SDT can be used for more authentic and efficient network evaluations than simulators and emulators. 

\subsection{Running Prevalent Network Functions on SDT}\label{sec:real_network_functions}

We also evaluate the feasibility of deploying prevalent network features on the SDT, with two specific modern network functions, RoCEv2, and a naive active routing. 

RoCEv2 works over lossless ethernet with PFC enabled. Since SDT does not have any hardware modifications to the physical ports, building a lossless network environment is possible by simply enabling the PFC on both switches and NIC ports. Moreover, DCQCN~\cite{zhu2015congestion} is an end-to-end congestion control method to delay the generation of PFC messages. Like PFC, the DCQCN can be enabled by directly turning it on as long as the switch and the NIC support it. We further deploy three types of deadlock avoidance methods alongside routing strategies on the SDT (Table~\ref{tab:routing-strategy}), which are working properly in the evaluation of real applications (See \S~\ref{sec:act_comparation}).

We implement an active routing algorithm based on~\cite{rahman2019topology} for the Dragonfly topology (a=4, g=9, h=2, with randomly selected 32 nodes). This algorithm extends Dragonfly's minimal routing policy by estimating network congestion according to the statistic data from \textit{Network Monitor} module. We evaluate active routing using a prevalent pure communication application, i.e., IMB Alltoall. Results show that active routing works well on SDT, which can reduce the ACT of the IMB Alltoall.

In summary, SDT shows strong adaptability to existing network functions. Most existing ethernet features can be easily deployed in SDT. Researchers can use SDT to validate existing network functions in multiple-scale topologies or to develop and evaluate new network functions using SDT.

\section{Discussion and Future Work}\label{sec:discussion}

\subsection{Flexibility Enhancement}

In SDT, the inter-switch links reservation issue might occur (\S~\ref{sec:multi-sdt}). Manual operations may still be required once the reserved inter-switch links cannot accommodate the new user-defined topology. To handle this, SDT can leverage optical switches to turn a link into either a self-link or an inter-switch link dynamically according to the topology requirements to further enhance the flexibility of SDT. We are designing the SDT controller with the optical switches and investigating whether there are additional challenges.

\subsection{Switch Selection}

The SDT controller in this paper performs TP operations on commodity OpenFlow switches. Generally, other switches can also be used for TP if they meet the following conditions: 1) allowing loopback packets to pass through self-links (or the STP protocol can be disabled), and 2) supporting 5-tuple matching or others similar to determine the forwarding of packets. For instance, other types of switches, like switches supporting extended ACL tables, are also suitable for TP. The P4-based (Intel Tofino) SDT controller is under refinement.

\subsection{Resource Limitation}

In SDT, the most significant resource is the maximum number of supported flow table entries in each OpenFlow switch. When a switch runs out of flow table entries during the setup of logical topology, the setup procedure may fail or other unknown failures could occur. SDT controller leverage a built-in module to check the number of available table entries to avoid such problem. If the demand for entries is greater than the available one, it can merge entries, split the topology, or inform operators to add more switches. In our evaluation, the problem of inadequate flow table capacity is rare. For instance, when we project a Fat-Tree with k=4 (containing 20 switches and 16 nodes) to 2 OpenFlow switches, each switch requires about only 300 flow table entries, which is not difficult for modern commercial OpenFlow switches to deploy.

\section{Conclusion}\label{sec:conclusion}

We summarize the advantages and disadvantages of existing network evaluation tools and conclude the methodology of an alternative method called ``Topology Projection'' (TP). Based on the idea of TP, we propose SDT, a deployment-friendly and automatically reconfigurable network topology testbed. SDT allows researchers to use several commodity OpenFlow switches to build network topologies based on user-defined topology configurations. SDT is fully transparent to other network components and can significantly reduce the deployment cost for network topology evaluations. We also develop the corresponding SDT controller for automatic topology reconfiguration. Through evaluations, we find that SDT can achieve almost the same physical properties as the full testbed and runs up to 2899x faster on network evaluations than the simulator does. SDT is more cost-effective and scalable than other TP solutions and can support a wide range of network research works.